\documentclass[review]{elsarticle}

\usepackage{lineno,hyperref}
\modulolinenumbers[5]

\journal{Journal of Simulation Modelling Practice and Theory}

\usepackage[ruled]{algorithm2e} 

\usepackage{lineno,hyperref}
\modulolinenumbers[5]

\usepackage{graphicx}
\usepackage{subfig}

\usepackage{multirow}
\usepackage{booktabs}
\usepackage{makecell}
\usepackage{tabularx}
\usepackage{ragged2e}

\usepackage{amsmath}
\usepackage{amsfonts}
\usepackage{euscript}
\usepackage{amssymb}

\usepackage{pdflscape}

\begin{document}

\begin{frontmatter}

\title{Modelling and Simulation Environment for Self-Adaptive and Self-Aware Cloud Architectures}

\author[mymainaddress]{Maria Salama \corref{mycorrespondingauthor}}
\cortext[mycorrespondingauthor]{Corresponding author}
\ead{m.salama@cs.bham.ac.uk}

\author[mymainaddress]{Rami Bahsoon}
\author[mysecondaryaddress]{Rajkumar Buyya}

\address[mymainaddress]{School of Computer Science, University of Birmingham, B15 2TT, UK}
\address[mysecondaryaddress]{Cloud Computing and Distributed Systems (CLOUDS) Lab, School of Computing and Information Systems, University of Melbourne, 3010 VIC, Australia}

\begin{abstract}
Cloud-based software systems are increasingly becoming complex and operating in highly dynamic environments. Self-adaptivity and self-awareness have recently emerged to cope with such level of dynamicity and scalability. Meanwhile, designing and testing such systems have poven to be a challenging task, as well as research benchmarking. Despite the influx of research in both self-adaptivity and cloud computing, as well as the various simulations environments proposed so far, there is a general lack of modelling and simulation environments of self-adaptive and self-aware cloud architectures. To aid researchers and practioners in overcoming such challenges, this paper presents a novel modelling and simulation environment for self-adaptive and self-aware cloud architectures. The environment provides significant benefits for designing self-adaptive and self-aware cloud architectures, as well as testing adaptation and awareness mechanisms. The toolkit is also beneficial as a symbiotic simulator during runtime to support adaptation decisions. We experimentally validated and evaluated the implementation using benchmarks and evaluation use cases.
\end{abstract}


\end{frontmatter}


\section{Introduction}
\label{sec_introduction}
Modern software systems are increasingly becoming large, complex, heterogeneous, pervasive, and tend to operate in unpredictable environments. Self-adaptivity has been motivated as a solution to achieve the level of dynamicity and scalability necessary for these systems, as well as to comply with the changes in components, fluctuations in workloads and environmental conditions during runtime \cite{Salehie2009} \cite{Cheng2009} \cite{Lemos2013}. Self-adaptive software architectures are expected to manage themselves following the principles of autonomic computing, to respond to changes in end-user requirments and the environment and to cope with uncertainty in runtime operation \cite{Oreizy1999} for continued satisfaction of quality requirements under changing context conditions \cite{Villegas2011}. Self-awareness has recently emerged to realise autonomic behaviour, with the aim of improving the quality of adaptation and seamlessly managing associated trade-offs \cite{Lewis2011} \cite{Faniyi2014}.

Adaptations decisions are taken during runtime with the aid of feedback loops (individual, collective or decentralised), analytical models, or by learning from historical data \cite{Salehie2009}. Symbiotic simulations are also powerful tools to support adaptation decisions. Such tools can be used symbiotically with the adaptation controller of the system, due to their ability to dynamically incorporate real-time data sensed from the system in running what-if scenarios and feedback the adaptation controller with the effects of adaptation decisions \cite{Aydt2009} \cite{Turner2011} \cite{Tjahjono2015} without causing extra overhead on the actual system.

Simulation-based approaches offer significant benefits to the research community and practitioners \cite{Quiroz2009} \cite{Calheiros2011}, supporting and accelerating research and development of systems, applications and services \cite{Buyya2009}. Simulaton tools are generally important and necessary software tools designed and developed to aid researchers, by allowing them to test their hypothesis or benchmarking studies in a controlled environment and easily reproduce results, perform experiments with different workloads and resource provisioning scenarios, as well as test systems performance \cite{Quiroz2009} \cite{Calheiros2011} \cite{Buyya2009}. In self-adaptive and self-aware software systems, simulation tools are needed to fill the gap between the conceptual research and the proof-of-concept implementation \cite{Nya2013}. Such tools help to systematically model and study the behaviour and performance of these systems that tend to operate in dynamically changing environments hard to define during system design \cite{Nya2013}. Simulations are also beneficial to self-adaptive and self-aware systems during runtime, as they can be used symbiotically with the adaptation controller of the system, where the results of the simulation are fed back to the system for taking adaptation decisions autonomously during runtime \cite{Aydt2009} \cite{Turner2011} \cite{Tjahjono2015}. In the context of cloud computing, simulators were known as tools to support and accelerate research and development of cloud computing systems, applications and services \cite{Buyya2009}, as quantifying the performance of service provision in real cloud environment is challenging \cite{Calheiros2011}. 

Given the highly dynamic operating environment of cloud computing and its on-demand nature \cite{Buyya2009} \cite{Armbrust2010}, cloud architectures tend to heavily leverage on adaptation to dynamically fulfil the uncertain and changing runtime demand \cite{Brandic2009} \cite{Antonopoulos2012} \cite{Maurer2013} \cite{Jamshidi2014} \cite{Chen2014}. The case of self-adaptive cloud architectures combines challenges of both clouds and self-adaptive architectures. In such case, testing architecture design or resources provisioning mechanisms, quantifying the architecture performance, and measuring the quality of service provisionned in real environments are challenging tasks. In the context of research, the benchmarking performance of a study under variable conditions and reproduction of results are difficult undertaking tasks \cite{Calheiros2011}. To this extent, we argue that simulation-based approaches are significantly important for research benchmarking, designing, testing and operating self-adaptive and self-aware cloud architectures. 

In this paper, we propose a novel modelling and simulation environment for self-adaptive and self-aware cloud architectures, namely \textit{SAd-CloudSim} and \textit{SAw-CloudSim}. The proposed toolkits build on the widely adopted cloud simulation environment \textit{CloudSim} \cite{Buyya2009} \cite{Calheiros2011}, due to its modular architecture that allows further extensions. CloudSim was found useful for this special case of cloud architectures, yet the former is not self-adaptive and self-aware by nature. The modelling and simulation environment can help in testing self-adaptive architectures design, understanding self-awareness behaviour, and evaluating associated trade-offs. The new extensions turn CloudSim to work with real systems at runtime as a symbiotic simulator, where self-adaptation and self-awareness helps in taking well-informed adaptation decisions. More specifically, \textit{SAd/SAw-CloudSim} provides an array of capabilities to model self-adaptation techniques and self-awareness capabilities. 

The \textit{SAd/SAw-CloudSim} tookits proposed in this paper offer the following novel extensions: (i) modelling and simulation of adaptation mechanisms and self-awareness for large-sale cloud-based systems, (ii) a self-contained platform for modelling and testing self-adaptation and self-awareness mechanisms, (iii) support for testing the performance of cloud systems under varying dynamic workloads and with different quality goals, (iv) the facility to simulate of architectural patterns with different combinations of self-awareness capabilities, and (v) support extensions for modelling and testing self-adaptation frameworks and self-awareness techniques. We validate and evaluateb the proposed toolkit with a series of experiments using the \textit{RUBiS} benchmark \cite{RUBiS} and the \textit{World Cup 1998} trend \cite{WorldCup98} and a number of evaluation use cases.

The rest of this paper is organised as follows. In section \ref{sec_background}, we describe relevant background and related work. Section \ref{sec_architecture} presents the architecture of the proposed framework. Section \ref{sec_design} presents technical details about the design and implementation. In section \ref{sec_evaluation}, we experimentally validate and evaluate the performance and overhead of our work. We discuss the threats to validity of the proposed work in section \ref{sec_threats}. Section \ref{sec_conclusion} concludes the paper and indicates future work.

\section{Background and Related Work}
\label{sec_background}
In this section, we discuss work related to simulators of self-adaptive, self-aware and cloud systems (section \ref{sec_relatedWork}). We, then, present background about self-adaptation and self-awareness (section \ref{sec_background_selfAdaptation}), as well as \textit{CloudSim} the cloud simulation toolkit on which we build our simulation environment (section \ref{sec_background_cloudSim}).

\subsection{Related Work}
\label{sec_relatedWork}
In the context of self-adaptive software systems, Abuseta et al. \cite{Abuseta2015} proposed a simulation environment for testing self-adaptive systems designed around the feedback control loop proposed by IBM architecture blueprint. A review for the state-of-the-art related to self-awareness in software engineering \cite{Elhabbash2018} has confirmed the lack of simulation tools for designing and evaluating such systems, with the exception of the work of \cite{Nya2013}. This work proposed a simulation environment for systems with self-aware and self-expressive capabilities, focusing on hardware aspects and precise process chronology execution. The simulation environment suits industrial relevant system sizes of avionic and space-flight industry.

With respect to cloud computing, there have been some notable proposals for simulation environments. An early survey has enlisted simulation approaches used for research in cloud computing \cite{Sakellari2013}. Examples include CloudSim \cite{Buyya2009} \cite{Calheiros2011} a modular and extensible open-source simulator, able to model very large scale clouds, GreenCloud \cite{Kliazovich2012} a packet-level simulator of energy-aware cloud data centers, MDCSim \cite{Lim2009} simulates multi-tier data centres in detail, and iCanCloud \cite{Nunez2012} \cite{Nunez2011} simulates cloud infrastructures flexibility and scalability. Other tools focused in simulating specific issues, such as power consumption and scientific worklflows \cite{Sakellari2013}. 

CloudSim has been widely adopted and used in many further extensions modelling and simulating cloud-related problems, due to its modular architecture. Examples include visually modelling and analysing cloud environments and applications (CloudAnalyst) \cite{Wickremasinghe2010}, modelling parallel applications (NetworkCloudSim) \cite{Garg2011}, simulating scientific worklflows (WorkflowSim) \cite{Chen2012}, concurrent and distributed cloud (Cloud2Sim) \cite{Kathiravelu2014a}, adaptive scaling cloud and MapReduce simulations (Cloud$^2$Sim) \cite{Kathiravelu2014b}, simulating heterogeneity in computational clouds (DynamicCloudSim) \cite{Bux2015}, and simulating containers in cloud data centres (ContainerCloudSim) \cite{Piraghaj2016}.

Despite the influx of research in self-adaptivity and cloud computing, as well as the various simulations environments proposed so far, there is a general lack, to the best of the authors knowledge, of modelling and simulation environments for self-adaptive and self-aware cloud architectures.

\subsection{Self-Adaptivity and Self-Awareness}
\label{sec_background_selfAdaptation}
Self-adaptivity is engineered to achieve the level of dynamicity and scalability necessary for modern and complex software systems, as well as to comply with the changes in components, fluctuations in workloads, and environmental conditions during runtime \cite{Salehie2009} \cite{Cheng2009}  \cite{Lemos2013}. A self-adaptive software ``evaluates its own behaviour and changes behaviour when the evaluation indicates that it is not accomplishing what the software is intended to do, or when better functionality or performance is possible'' \cite{Laddaga1997} \cite{Oreizy1999} \cite{Cheng2009a}. Intuitively, a self-adaptive system is one that has the capability of modifying its behaviour at runtime in response to changes in the dynamics of the environment (e.g. workload) and disturbances to achieve its goals (e.g. quality requirements) \cite{Meng2001}. Self-adaptive systems are composed of two sub-systems: (i) the managed system (i.e. the system to be controlled), and (ii) the adaptation controller (the managing system) \cite{Villegas2011}. The managed system structure could be either a non-modifiable structure or modifiable structure with/without reflection capabilities (e.g. reconfigurable software components architecture) \cite{Villegas2011}. The controller's structure is a variation of the MAPE-K loop \cite{Villegas2011}.

Self-adaptive architectures are expected to manage themselves following the principles of autonomic computing, to respond to environmental changes and prevent service provision violations \cite{Oreizy1999}. Examples of adaptation strategies include architectural tactics, as mechanisms for better tuning, responding and achieving Quality of Service (QoS) attributes, such as response time, throughput, energy efficiency. Architectural tactics are inherently architectural decisions, with measurable response, designed to support quality attributes subject of interest \cite{Bass2003} \cite{Mirakhorli2012}. For instance, tactics are designed for performance, greenability, availability, and reliability, e.g. horizontal scaling, vertical scaling and VMs consolidation \cite{Bass2003} \cite{Procaccianti2014}.

As self-adaptive software systems are increasingly becoming heterogeneous with dynamic requirements and complex trade-offs \cite{Nya2014}, engineering self-awareness and self-expression is an emerging trend in the design and operation of these systems. Inspired from psychology and cognitive science, the concept of self-awareness has been re-deduced in the context of software engineering to realise autonomic behaviour for software exhibiting these characteristics, with the aim of improving the quality of adaptation and seamlessly managing these trade-offs \cite{Lewis2011} \cite{Faniyi2014}.

The self-aware architecture style draws on the principles of self-awareness to enrich self-adaptive architectures with self-awareness capabilities. As the architectures of such software exhibit complex trade-offs across multiple dimensions emerging internally and externally from the uncertainty of the operation environment, the self-aware architecture style is designed in a fashion where adaptation and execution strategies for these concerns are dynamically analysed and managed at runtime. 

Self-aware architecture style is defined based on a \textit{self-aware node} unit \cite{Faniyi2014}. A self-aware computational node is defined as a node that ``possesses information about its internal state and has sufficient knowledge of its environment to determine how it is perceived by other parts of the system'' \cite{Lewis2011} \cite{Faniyi2014}. A node is said to have \textit{self-expression} capability ``if it is able to assert its behaviours upon either itself or other nodes, this behaviour is based upon a nodes sense of its personality'' \cite{Parsons2011}\footnote{Architecting self-aware software has been introduced in \cite{Faniyi2014} and detailed in \cite{Chen2014a}}. Different levels of self-awareness, called capabilities, were identified to better assist the self-adaptive process \cite{Parsons2011} \cite{Faniyi2014}: 

\begin{itemize}
	\item \textit{Stimulus-awareness}: a computing node is stimulus-aware when having knowledge of stimuli, enabling the system's ability to adapt to events. This level is a prerequisite for all other levels of self-awareness. 
	\item \textit{Goal-awareness}: if having knowledge of current goals, objectives, preferences and constraints, in such a way that it can reason about it. 
	\item \textit{Interaction-awareness}: when the node's own actions form part of interactions with other nodes and the environment. 
	\item \textit{Time-awareness}: when having knowledge of historical information and/or future phenomena. 
	\item \textit{Meta-self-awareness}: the most advanced of the self-awareness levels, which is awareness of own self-awareness capabilities.
\end{itemize}

\begin{figure*}[!h]
\centering
\includegraphics[width=0.85\textwidth]{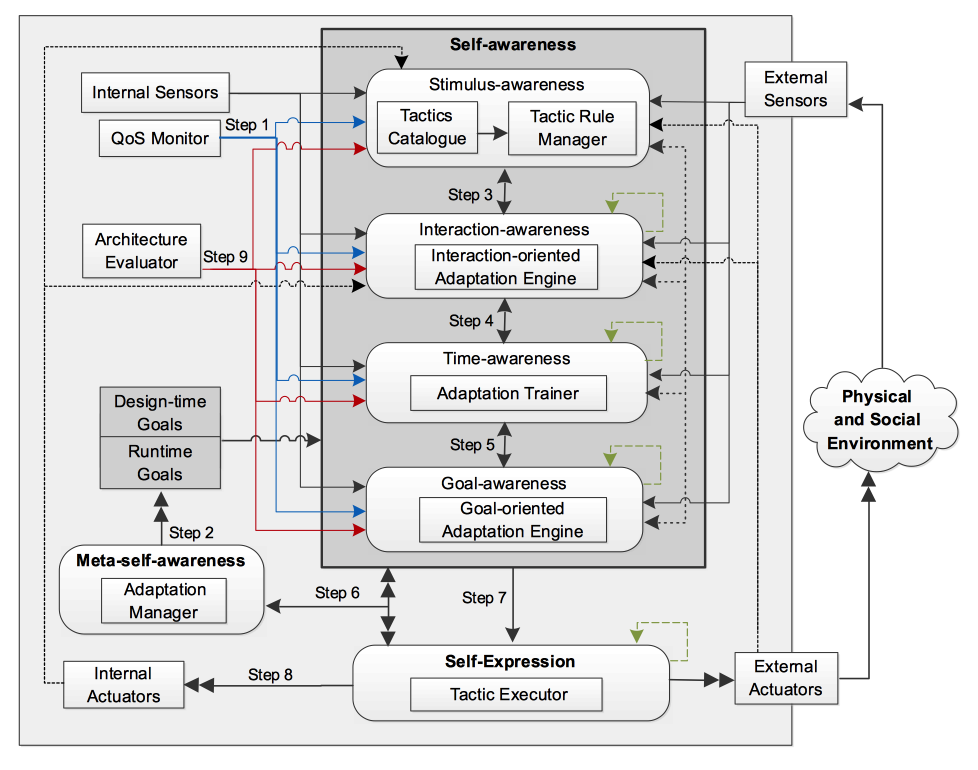}
\caption{Quality-driven self-aware architecture pattern \cite{Salama2015}}
\label{fig_metaPattern}
\end{figure*}

Various architecture patterns were introduced, each contains different self-awareness capabilities, so that the pattern used when designing the software, would include capabilities relevant to the software requirements. Further, self-aware architectural patterns have been enriched with quality self-management capabilities, in order to achieve the desired quality of service levels in a seamless way \cite{Salama2015}. Figure \ref{fig_metaPattern} shows the architecture pattern featuring all levels of awareness with the tactics generic components. Architectural patterns have been enriched with a catalogue of architectural tactics designated to fulfil different quality attributes. Incorporating the tactics, as adaptation actions to meet the quality requirements, aims at improving and enriching the quality of self-expression, i.e. the adaptation actions taken by the self-aware node. The selection of the appropriate tactic is performed during runtime by the awareness capabilities available at different patterns. 

\subsection{CloudSim}
\label{sec_background_cloudSim}
The \textit{CloudSim} simulation toolkit \cite{Calheiros2011} \cite{Buyya2009} is currently one of the mostly-used general purpose cloud simulation environments \cite{Kathiravelu2014a}, and the most sophisticated discrete event simulator for clouds \cite{Garg2011}. Due to its modular architecture, it has been widely adopted and used in many further extensions modelling and simulating cloud-related problems.

\begin{figure}[!h]
\centering
\includegraphics[width=0.65\textwidth]{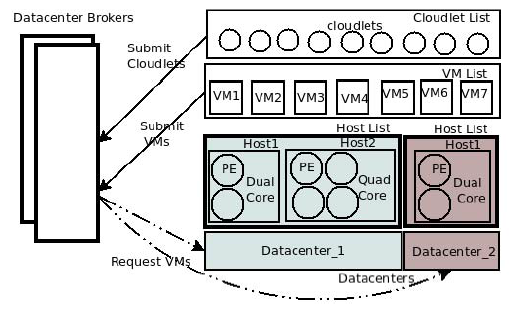}
\caption{CloudSim Architecture \cite{Calheiros2011}}
\label{fig_cloudsimArchitecture}
\end{figure}

Figure \ref{fig_cloudsimArchitecture} shows a cloud environment represented by the architecture of CloudSim. CloudSim defines the core entities of a cloud environment, such as datacenters, hosts physical machines (PMs), virtual machines (VMs), applications or user requests (called cloudlets) \cite{Buyya2009} \cite{Calheiros2011}. Datacenter is the resources provider, simulating the infrastructure of the cloud, and hosts which run virtual machines responsible for processing user requests. Computational capacities of PMs and VMs (CPU unit) are defined by $ Pe $ (Processing Element) in terms of million instructions per second (MIPS) \cite{Buyya2009} \cite{Calheiros2011}. Processing elements in a PM are shared among VMs, and among requests in a VM. The Datacenter Broker is responsible about the allocation of requests to VMs. Once the simulation period is started, the requests are scheduled for execution, and the cloud behaviour is simulated. 

\section{SAd/SAw-CloudSim Architecture}
\label{sec_architecture}
\begin{figure}[!h]
\centering
\includegraphics[width=0.65\textwidth]{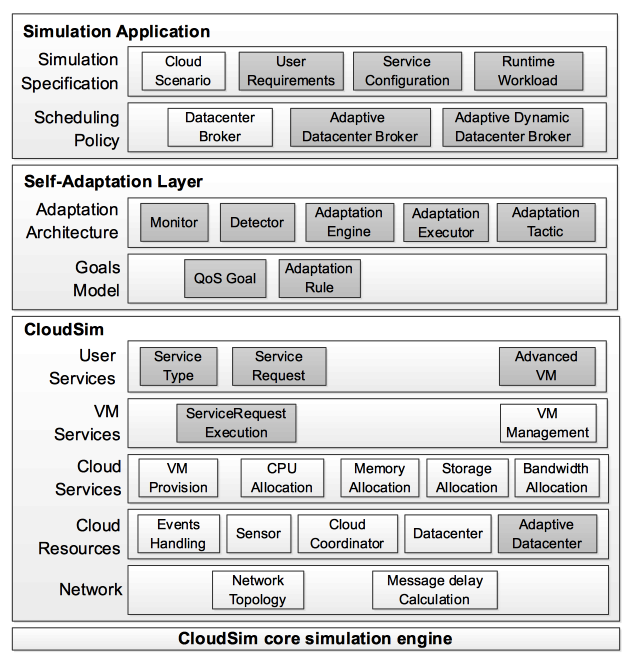}
\caption{\textit{SAd-CloudSim} architecture}
\label{fig_adcloudsimArchitecture}
\end{figure}

\begin{figure}[!h]
\centering
\includegraphics[width=0.65\textwidth]{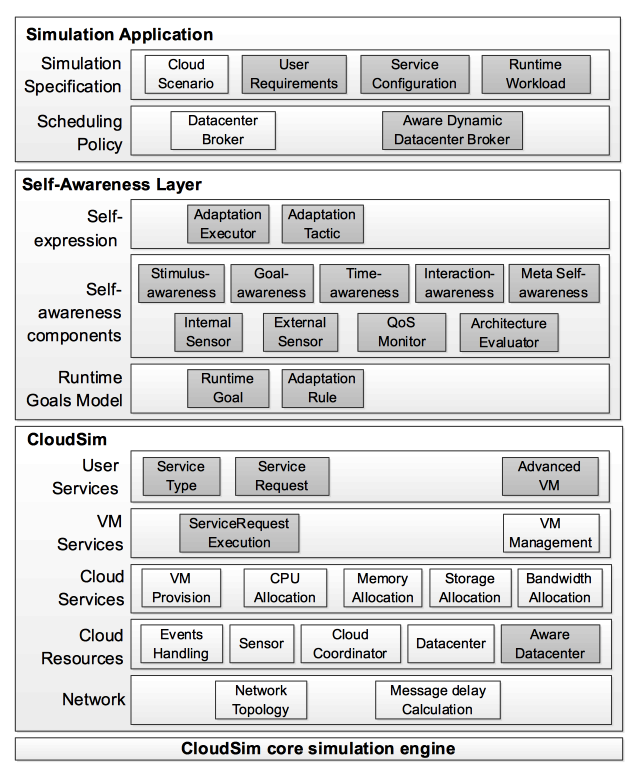}
\caption{\textit{SAw-CloudSim} architecture}
\label{fig_awcloudsimArchitecture}
\end{figure}

In this section, we outline the architecture of SAd/SAw-CloudSim, the extensions made to CloudSim core framework and the rationale behind them. Figures \ref{fig_adcloudsimArchitecture} and \ref{fig_awcloudsimArchitecture} show the multi-layered design of CloudSim with the architectural components of \textit{SAd-CloudSim} and \textit{SAw-CloudSim} respectively (new components are shown in dark boxes).  

Generally, the proposed environments are built on top of the CloudSim core simulation engine and CloudSim core. Extensions for some core classes of CloudSim were necessary for adaptation and awareness capabilities (more details in section \ref{sec_design}). The Self-Adaptation layer is added on top of the cloud core architecture, to model the adaptation controller of a self-adaptive software system. Researchers and practitioners, willing to design an adaptation technique or study the efficiency of an existing one, would need to implement their techniques in this layer. The Self-Awareness layer combines the self-awareness and self-expression capabilities, as well as necessary monitoring components. The top-most layer is the Simulation Application, inherited from CloudSim, that models the specification of the simulation to be conducted using the tool. Such specifications allow to configure the simulation of dynamic workloads, different service types and user requirements.

\subsection{Modelling Self-Adaptation}
A foundational self-adaptation controller consists of: (i) monitor for correlating quality data, (ii) detector for analysing the data provided by the monitor and detecting violations in order to trigger adaptation when necessary, (iii) adaptation engine to determine what needs to be changed and select the optimal adaptation strategy, and (iv) adaptation executor responsible for applying the adaptation action on the underlying infrastructure. Our initial implementation of \textit{SAd-CloudSim} includes this foundational version of adaptation controller. Such components could be further extended to study more complex adaptation mechanisms, such as pro-active adaptations or MAPE-K adaptation process \cite{Salehie2009}.

The \textit{Monitor} component is responsible for monitoring the achievement of quality requirements. The \textit{Detector} checks any violations occurring during runtime against quality goals. Whenever a violation is detected, adaptation is triggered. The \textit{Adaptation Engine} is responsible for analysing the current situation and selecting the optimal adaptation strategy that would achieve the quality targets, e.g. increasing VMs capacity, increase number of PMs. The selected adaptation tactic is executed dynamically during runtime on the cloud infrastructure by the \textit{Adaptation Executor}. 

\subsection{Modelling Self-Awareness}
Modelling self-awareness capabilities in a cloud architecture requires the following components: (i) QoS monitoring, (ii) different self-awareness capabilities as the system requires, and (iii) self-expression capability to execute adaptations. The monitoring component is composed of sensors responsible for measuring actual quality data, and the QoS monitor responsible for correlating data from sensors and monitoring changes in workload and quality attributes during runtime. The self-awareness component contains different awareness capabilities enabled according to the system requirements. The stimulus-awareness is the basic awareness capability responsible for triggering adaptations when a violation is detected and selecting an adaptation tactic from the tactics catalogue. Other self-awareness capabilities help in selecting the optimal tactic using their owned information. For instance, time-awareness can provide historical information about the performance of a tactic under similar conditions. The goal-awareness is capable to detect possible violations within a threshold. The meta-self-awareness decides on which awareness level the architecture would operate. The selected tactic is executed by the \textit{Adaptation Executor} of the self-expression component. The \textit{Architecture Evaluator} evaluates the new state after executing the tactic, where such information is passed to the time-awareness component.

\subsection{Modelling QoS Goals and Adaptation Tactics}
Goals are the main objective or trigger for self-adaptation. \textit{QoS Goals} represent the quality of service targets required to be fulfilled. Whenever violated, an adaptation should take place to achieve the quality goals. For each QoS Goal, a set of possible adaptation tactics is implemented in the tactics catalogue. Also, adaptation rules are defined as \textit{if-condition-then-action} rules, where the conditions are quality requirements and the actions are response tactics.

In \textit{SAd-CloudSim}, the \textit{Goals Model} combines these quality targets. For self-adaptive architectures, goals are specified as static values for quality attributes required to be fulfilled. These values are checked during runtime against the actual quality measured data, and adaptations are triggered whenever a violation is detected.

Employing self-awareness capabilities requires a more sophisticated goals model, where \textit{Runtime Goals} can be dynamically settled at runtime or specified for different users. The \textit{Runtime Goals Model} keeps historical information about the satisfaction of goals and the performance of adaptation tactics to be used for better informed decision when choosing the optimal tactic and for future learning using the time-awareness capability.

\section{Design and Implementation}
\label{sec_design}
In this section, we provide details related to the classes and implementation of \textit{SAd-CloudSim} and \textit{SAw-CloudSim}.

\subsection{Extensions to CloudSim Core}
We have extended some core classes of CloudSim by adding necessary quality and power (energy) metrics, namely AdaptiveDatacenter, AwareDatacenter, AdvancedHost and AdvancedVM. The DatacenterBroker \textemdash responsible for workload distribution and resources provisioning \textemdash is also extended by queueing models necessary for adaptation and awareness capabilities. A \textit{RuntimeWorkload} is added to allow conducting experiments for consecutive time intervals, and user requirements are added to configure QoS requirements.

We use the \textit{Service Type} class to model an SaaS service offered by the cloud provider. A service type is configured by the computational resources it requires (MIPS). A \textit{Service Request} is used to model a request made by an end-user for a specific service type. This allows modelling dynamic workloads by multiple end-users for a variety of services. 

\subsection{Self-Adaptation Simulation}
\begin{figure}[!h]
\centering
\includegraphics[width=0.65\textwidth]{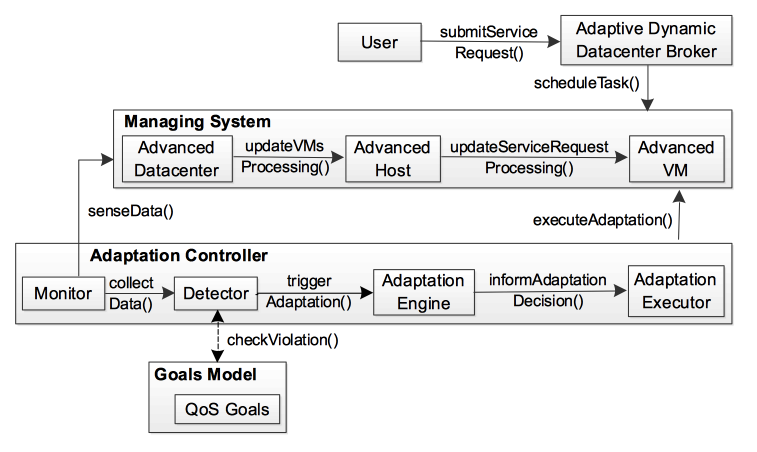}
\caption{Self-Adaptation simulation process}
\label{fig_adaptationSimProcess}
\end{figure}

The \textit{Self-Adaptation} package encapsulates the components necessary for modelling and simulating a self-adaptive architecture. Our initial implementation includes the basic functionalities of these components. Figure \ref{fig_adaptationSimProcess} depicts the flow of the simulation process in case of self-adaptation. These components could be further extended with more sophisticated implementations, such as MAPE-K. This package is composed of the following classes:

\begin{itemize}
\item \textit{Self-Adaptive Architecture} class is the main class responsible for instantiating and managing the adaptation components, i.e. monitor, detector, adaptation engine, adaptation executor. Once instantiated, it loads the goals model from the user configuration \texttt{xml} file. It is also responsible for keeping track of the adaptation history and overhead for performance evaluation. This class is designed using the singleton pattern.

\item \textit{Goals Model} class is the list of goals objects loaded from a configuration file. Each \textit{Goal} object contains the list of attributes, that are: goal id, name, constraint value, metric (e.g. ms), objective (if the objective is to minimise or maximise the attribute), weight, a boolean indicator whether it is violated. The constraint value is the requirement to be achieved.

\item \textit{Monitor} class runs as a thread in the background. It contains methods sensing, measuring and collecting actual data of the QoS parameters of the executed requests, e.g response time, throughput, energy consumption. The monitor is configured with the monitoring frequency to run and collect data. After cleaning the queue of the previous monitoring cycle, the collected data is put in the queue to be sent to the detector.

\item \textit{Detector} class contains a method triggered to run after receiving data from the monitor. It checks the runtime values of the quality metrics against the Goals Model. If a violation is detected, adaptation is triggered.

\item \textit{Adaptation Engine} class is responsible for selecting the optimal adaptation action after receiving the adaptation trigger. The adaptation action is selected from the Adaptation Tactics Catalogue according to the adaptation rules. Adaptations rules list object is set in this class using \texttt{xml} configuration file that contains the quality attributes, their associated tactics and their order of execution. The selection is based on simple rule-based algorithm, and could be further extended with knowledge-based models.

\item \textit{Adaptation Tactics Catalogue} class contains a list of adaptation tactics, loaded from \texttt{xml} configuration file. Examples of tactics could be increasing VMs capacity, number of VMs or PMs for better response time and consolidating VMs for less energy consumption. Each \textit{Adaptation Tactic} object contains  the attributes of a tactic, that are: id, description, affected object (e.g., host, VM), change (increase or decrease) and the minimum and maximum limits (e.g. minimum one running host and maximum capacity of the datacenter).

\item \textit{Adaptation Rule} class links quality attributes with their adaptation tactics. It contains the details of an adaptation rule, that are: id, description, quality attribute, adaptation tactic and its priority in execution.

\item \textit{Adaptation Executor} class performs the actual execution of the selected adaptation action on the relevant object, i.e. VM instances, list of VMs, list of PMs.
\end{itemize}

\subsection{Self-Awareness Simulation}
\begin{figure}[!h]
\centering
\includegraphics[width=0.65\textwidth]{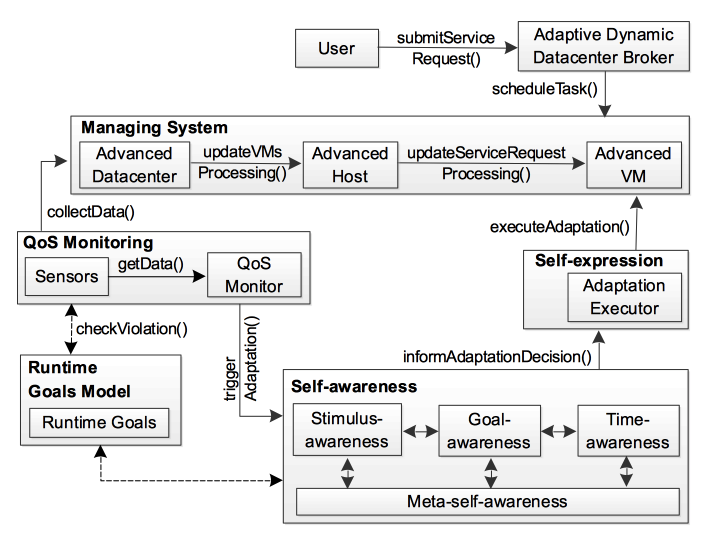}
\caption{Self-Awareness simulation process}
\label{fig_awarenessSimProcess}
\end{figure}

Figure \ref{fig_awarenessSimProcess} depicts the flow of the simulation process in case of employing self-awareness and self-expression capabilities. The \textit{Self-Awareness} package encapsulates the components necessary for modelling and simulating a self-aware and self-expressive architecture, as follows.

\begin{itemize}
\item \textit{Self-aware Architecture} class is the main class responsible for instantiating and managing the main components, i.e. QoS Monitoring, Self-Awareness and Self-Expression components. Once instantiated, it loads the runtime goals model from the user configuration \texttt{xml} file. Its also responsible for keeping track of the adaptation history and overhead for performance evaluation. This class is designed using the singleton pattern.

\item \textit{Runtime Goals Model} class contains the list of runtime goals objects loaded from the configuration file. Each \textit{Runtime Goal} object is inherited from the Goals Model class, and contains new set of attributes: user id (to mark the runtime goals of different users) and violation threshold (to reflect the threshold to take pro-active adaptations). The Runtime Goal Model contains history records to keep track of the goals fulfilment (i.e. time instance, average violation value, tactic executed, average value after adaptation).

\item \textit{QoS Monitoring} component is composed of sensors for different quality requirements, QoS Monitor and Architecture Evaluator, as described below:
\begin{itemize}
\item \textit{Internal Sensor} and \textit{External Sensor} classes contain methods running in the background for continuously sensing data about QoS parameters. The internal sensors are for sensing the actual quality parameters in the self-aware node. The external sensors are required for interaction-awareness for sensing data from the other nodes with which the node is interacting. 

\item \textit{QoS Monitor} class contains another background method is for correlating data received from the sensors. Such data is sent to the self-awareness component to take necessary actions. The basic version of the QoS Monitor constantly send data to the self-awareness component. More sensitive monitors can vary the interval of data correlation according to sensed data.

\item \textit{Architecture Evaluator} class continuously evaluates the response after executing the adaptation action and feeds the different levels of awareness for further actions if needed.
\end{itemize}

\item \textit{Self-Awareness} component encompasses the different levels of self-awareness. These levels that could be enabled as per the relevance to the system requirements using a configuration file. Each self-awareness component is designed using the \textit{Self-Awareness} abstract class to implement the \texttt{act} method. Self-awareness components are: 
\begin{itemize}
\item \textit{Stimulus-awareness} class embeds rules for selecting and composing optimal adaptation actions or tactics, by defining ``if-condition-then-action'' rules where the conditions are quality parameters subject of violation and actions are response tactics. The adaptation is triggered when violations are detected.

\item \textit{Goal-awareness} class contains the \texttt{act} method operating as a ``goal-oriented adaptation engine'' that uses knowledge about runtime goals to make decisions about the tactic selection in line with the systems goals. This version of adaptation engine is more sensitive towards violations, and can take pro-active actions before violations.

\item \textit{Time-awareness} class contains the adaptation trainer method that uses historical data about tactics responses under different runtime conditions to improve the quality of adaptation. Implementing machine learning techniques is useful for realising time-awareness.

\item \textit{Interaction-awareness} class contains the interaction-oriented adaptation engine that should contribute to the selection of the tactic according to the runtime environmental conditions of other nodes. This, currently implemented as an abstract, could be implemented in cases of distributed clouds or cloud federations \footnote{currently beyond the scope of this work}.

\item \textit{Meta-self-awareness} class contains the adaptation manager method to reason about the benefits and costs of maintaining a certain level of awareness (and degree of complexity with which it exercises this level), as well as the benefits and costs of selecting a tactic based on a certain level of awareness. This can also dynamically select a particular adaptation out of a set of possibilities for realising one or more levels, in order to manage trade-offs between different QoS attributes. Trade-offs management algorithms could be implemented here. A more sophisticated \texttt{act} method can adapt the way in which the level(s) of self-awareness are realised, e.g. by changing algorithms realising the level(s), thus changing the degree of complexity of realisation of the level(s).  
\end{itemize}

\item \textit{Self-expression} component is responsible for the execution of the adaptation decision made by the self-awareness component. It is composed of the \textit{Adaptation Executor} responsible for managing the process of adaptation execution during runtime. In more details, it makes necessary instructions about the composition and instantiation of the components required for the adaptation decision. As an example, in the case of VMs consolidation, it decides which VMs should be consolidated, where these VMs should be placed, which PMs should be switched off. Then, it performs the actual instantiation of the tactic components during runtime, such as creating new VMs or switching off PMs.

\end{itemize}

\section{Experimental Validation and Evaluation}
\label{sec_evaluation}
This section aims to examine the capability of the proposed framework to instantiate different architectures of cloud nodes, validate the self-adaptation and self-awareness components, and assess associated overhead. In the course of the validation process, we do not contribute with new scheduling policies. We use current scheduling policies to test the new simulation toolkits.

\subsection{Architecture Instantiation}
We instantiated the architecture of a cloud node using the self-adaptive and quality-driven self-aware pattern to perform QoS-driven adaptations. To this end, this architecture dynamically performs architecture-based adaptation, which uses the knowledge available in choosing optimal tactics to meet QoS requirements during runtime.

The QoS attributes, to be taken into consideration in this case (as defined in \cite{Chen2014a}, \cite{Salama2017}), include: (i) quality requirements specified in end-users SLAs, (ii) environmental restrictions, and (iii) economic constraints. Table \ref{tbl_QoSAttributes} lists details of the QoS attributes. With respect to the quality requirements, we consider performance (measured by response time from the time the user submits the request till the cloud submits the response back to the user in milliseconds). For the environmental aspect, we use the greenability property \cite{Lago2015} \cite{Calero2015} measured by energy consumption in kWh. For the economic constraints, we define the operational cost by the cost of computational resources (CPUs, memory, storage and bandwidth). The goals objectives are defined to be challenging.

We define the catalogue of architectural tactics to fulfil the quality attributes subject to consideration. Table \ref{tbl_QualityTactics} lists the tactics and their definitions. We base this work on the description tactics by Bass et al. \cite{Bass2012}. The tactics include: (i) horizontal scaling (increasing/decreasing the number of physical machines), (ii) vertical scaling (increasing/decreasing the number of virtual machines or their CPU capacities), (iii) virtual machines consolidation (running the virtual machines on less number of physical machines for energy savings), (iv) concurrency (by processing different streams of events on different threads or by creating additional threads to process different sets of activities), (v) dynamic priority scheduling (scheduling policy is implemented, where the scheduler handles requests according to a scheduling policy), and (vi) energy monitoring (providing detailed energy consumption information). Adaptation rules are, then, embedded in the adaptation engine and the stimulus-awareness component, where tactics are related with the QoS attributes. Adaptation rules are illustrated in Table \ref{tbl_AdaptationRules}.

\begin{table}[!h]
\caption{QoS attributes}
\label{tbl_QoSAttributes}
\center
\footnotesize
\begin{tabular}
{
>{\raggedright}p{0.20\textwidth} 
>{\raggedright}p{0.10\textwidth} 
>{\raggedright}p{0.10\textwidth} 
>{\raggedright\arraybackslash}p{0.10\textwidth} 
}
\toprule
	\textbf{Attribute}						&
	\textbf{Weight}							&
	\textbf{Metric}								&	
	\textbf{Objective}				
	\\
	\midrule
Response time							&
0.50										&
ms											&
25	
\\
Greenability							&
0.20										&
kWh										&
25 
\\
Operational cost					&
0.20										&
\$												&
50		
\\
	\bottomrule
\end{tabular}
\end{table}

\begin{table*}[!h]
\caption{QoS tactics and their definitions}
\label{tbl_QualityTactics}
\center
\footnotesize
\begin{tabular}
{
l
>{\raggedright}p{0.12\textwidth} 
>{\raggedright}p{0.25\textwidth} 
>{\raggedright}p{0.10\textwidth} 
>{\raggedright}p{0.15\textwidth} 
>{\raggedright\arraybackslash}p{0.22\textwidth} 
}
	\toprule
	\textbf{No.}						&
	\textbf{Tactic}					&	
	\textbf{Description}		&	
	\textbf{Object}				&
	\textbf{Limits}					&	
	\textbf{Variations}			
	\\
	\midrule
1											&
Vertical scaling 				&
increasing the number of virtual machines (VMs) or their CPU capacities							&
VMs									&
maximum CPU capacity of hosts running in the datacenter		&
+1, 2, 3,... VMs or increase the CPU capacity of running VMs	
\\
2											&
Vertical de-scaling 				&
decreasing the number of virtual machines (VMs) or their CPU capacities									&
VMs											&
minimum one running VM		&
+1, 2, 3,... VMs						
\\
3											&
Horizontal scaling			&
increasing the number of running hosts					&
Hosts									&
maximum number of hosts in the datacenter			&
+1, 2, 3,... hosts				
\\
4											&
Horizontal de-scaling					&
decreasing the number of running hosts					&
Hosts													&
minimum one running host 			&
-1, 2, 3,... hosts								
\\
5											&
VMs consolidation			&
shut down hosts running least number of VMs and migrate their VMs to other hosts									&
Hosts, VMs												&
minimum one running host and one VM 			&	 
-1, 2, 3,... hosts										
\\
6															&
Concurrency 									&
processing different streams of events on different threads or by creating additional threads to process different sets of activities		&
datacenter scheduler					&
maximum CPU capacity of hosts running in the datacenter					&
single, multiple threads					
\\
7												&
Dynamic scheduling			&
scheduling policy is implemented, where the scheduler handles requests according to a scheduling policy						&
datacenter scheduler						&
maximum number of running hosts and VMs					&
earliest deadline first scheduling, least slack time scheduling, single queueing, multiple queueing, multiple dynamic queueing				
\\
\bottomrule
\end{tabular}
\end{table*}

\begin{table}[!h]
\caption{Adaptation Rules}
\label{tbl_AdaptationRules}
\center
\footnotesize
\begin{tabular}
	{ l 
	l  
	l }
	\toprule
	\textbf{Tactic}												&	
	\textbf{Related Quality Attributes}		&
	\textbf{Priority}
	\\
	\midrule
Dynamic scheduling						&
response time 									&
1
\\
Concurrency 									&
response time									&
2
\\
Vertical scaling 									&
response time										&
3
\\
Horizontal scaling							&
response time									&
4
\\
VMs consolidation												&
operational cost, energy consumption			&
1
\\
Vertical de-scaling 												&
operational cost, energy consumption				&
2
\\
Horizontal de-scaling										&
operational cost, energy consumption			&
3
\\
	\bottomrule
\end{tabular}
\end{table}

\begin{figure*}[!t]
\centering
\includegraphics[width=0.90\textwidth]{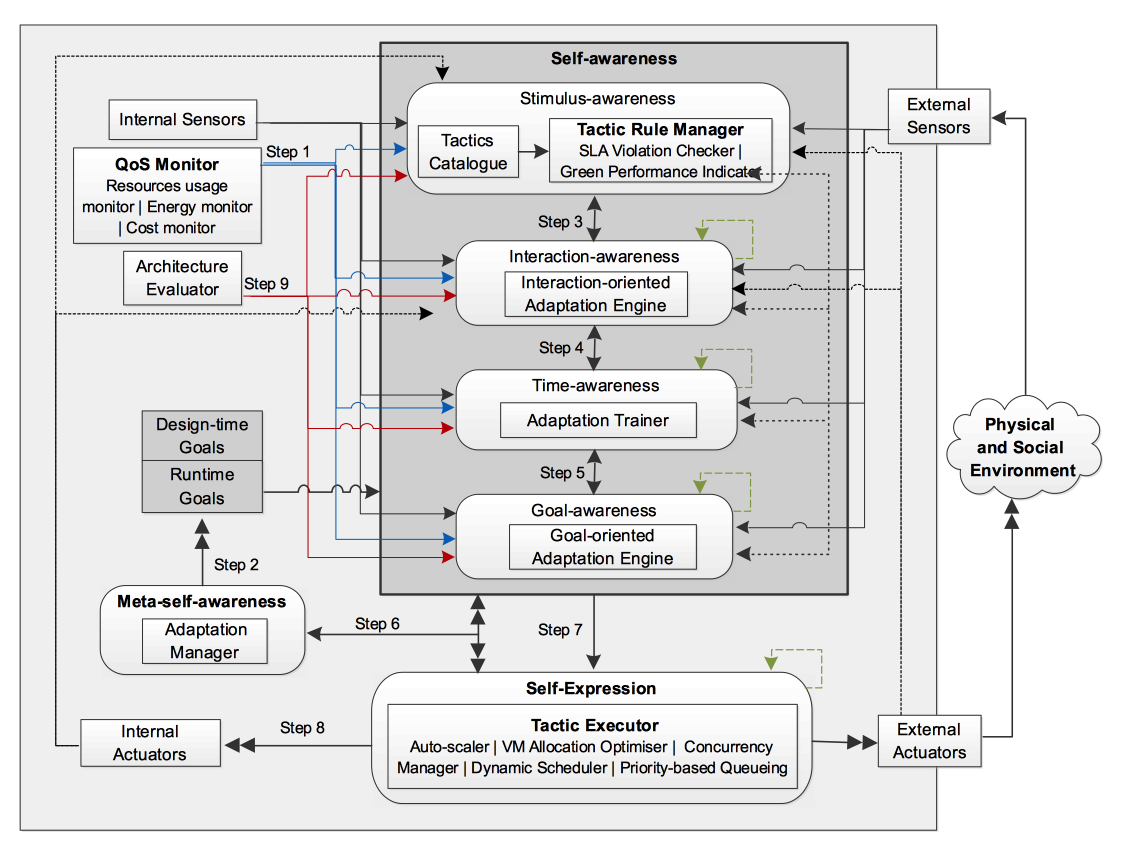}
\caption{Cloud architecture instantiated using quality-driven self-aware pattern}
\label{fig_cloudArchitecture}
\end{figure*}

We embed the tactics catalogue in the self-adaptive and self-aware architectures and the relationships are made implicit within the interaction between different components. The architecture of the self-aware cloud node is illustrated in Figure \ref{fig_cloudArchitecture}. Tactics are defined in the Tactics Catalogue component. Monitors for quality attributes are implemented in the QoS Monitor component. Components necessary for checking possible violation of quality attributes are implemented in the stimulus-awareness component, e.g. SLA Violation Checker and Green Performance Indicator. The scheduler component of the scheduling tactic was embedded into the stimulus-aware. Management components of tactics were configured into the Tactic Executor for running the tactics, e.g. auto-scaler. 

\subsection{Testbed Configuration}
We used benchmarks to stress the architecture with highly frequent changing demand and observe quality goals. To simulate runtime dynamics, we used the \textit{RUBiS} benchmark \cite{RUBiS} and the \textit{World Cup 1998} trend \cite{WorldCup98} in our experiments. The \textit{RUBiS} benchmark \cite{RUBiS} is an online auction application defining different services categorised in two workload patterns: the browsing pattern (read-only services, e.g. BrowseCategories), and the bidding pattern (read and write intensive services, e.g. PutBid, RegisterItem, RegisterUser). For fitting the simulation parameters, we mapped the different services of the \textit{RUBiS} benchmark into Million Instructions Per Second (MIPS), as listed in Table \ref{tbl_experimentsServices}. To simulate a realistic workload within the capacity of our testbed, we varied the number of requests proportionally according to the \textit{World Cup 1998} workload trend \cite{WorldCup98}. We compressed the trend in a way that the fluctuation of one day (= 86400 sec) in the trend corresponds to one time instance of 864 seconds in our experiments. This setup can generate up to 700 parallel requests during one time instance, which is large enough to challenge quality. 

\begin{table}[!h]
\caption{Types of service requests}
\label{tbl_experimentsServices}
\center
\footnotesize
\begin{tabular}
{
		>{\raggedright}p{0.25\textwidth} 
		l
		l
		l
}
\toprule
{\textbf{Service Pattern}} 		& 
{\textbf{S\#}} 								& 
{\textbf{Service Type}} 			& 
{\textbf{Required MIPS}}
\\
\midrule
browsing only			&
1									&
read-only					&
10,000  
\\
\hline
bidding only				&
2									&
read and write			&
20,000  
\\
\hline
\multirow{3}{*}{\parbox{3cm}{mixed with adjustable composition of the two service patterns}}															&
3																											&
70\% browsing, 30\% bidding												&
12,000  
\\
																											&
4																											&
50\% browsing, 50\% bidding												&
15,000  
\\
																											&
5																											&
30\% browsing, 70\% bidding												&
17,000  
\\
\bottomrule
\end{tabular}
\end{table}

The configuration of the datacenter hosts is IBM x3550 server of 2 x Xeon X5675 3067 MHz, 6 cores and 256 GB RAM. The frequency of the servers' CPUs are mapped onto MIPS ratings: 3067 MIPS each core \cite{Beloglazov2012} and their energy consumption is calculated using power models of \cite{Beloglazov2012}. The maximum capacity of the cloud datacenter is 1000 hosts. The characteristics of the virtual machines (VMs) types correspond to the latest generation of General Purpose Amazon EC2 Instances \cite{AmazonEC2}. In particular, we use the \texttt{m4.large} (2 core vCPU 2.4 GHz, 8 GB RAM), \texttt{m4.xlarge} (4 core vCPU 2.4 GHz, 16 GB RAM) and \texttt{m4.2xlarge} (8 core vCPU 2.4 GHz, 32 GB RAM) instances. The operational cost of different VMs types is 0.1, 0.2 and 0.4 \$/hour respectively. The initial deployment of the experiments is shown in Table \ref{tbl_experimentsConfiguration}. When running self-adaptive, stimulus-aware and goal-aware architectures, the initial deployment is 10 hosts running 15 VMs. Initially, the VMs are allocated according to the resource requirements of the VM types. However, VMs utilise less resources according to the workload data during runtime, creating opportunities for dynamic consolidation. For the non-adaptive architecture, the deployment is 70 hosts running 210 VMs (the maximum number used by the self-adaptive architecture) to allow processing the maximum number of requests during peak load.

\begin{table}[!h]
\caption{Initial deployments of the experiments}
\label{tbl_experimentsConfiguration}
\center
\footnotesize
\begin{tabular}
{
	l
	>{\raggedright\arraybackslash}p{0.45\textwidth} 
}
\toprule
{\textbf{Configuration}} 		& 	
{\textbf{}}					
\\
\midrule
Hosts type		&
IBM x3550 server
\\
Hosts Specs	&
2 x Xeon X5675 3067 MHz, \newline 6 cores, 256 GB RAM
\\
\hline
VMs types		&
General Purpose Amazon EC2 Instances 
\\
VMs Specs			&
m4.large: 2 core CPU 8 GB RAM	\newline
m4.xlarge: 4 core CPU 16 GB RAM 	\newline
m4.2xlarge: 8 core CPU 32 GB RAM
\\
\hline
No. of hosts			&
non-adaptive: 70			\newline			
adaptive: 10 (max. 1000)		
\\
No. of VMs						&	
non-adaptive: 210 x m4.xlarge		\newline
adaptive: \newline 5 x m4.large,	 5 x m4.xlarge,	 5 x m4.2xlarge
\\
\bottomrule
\end{tabular}
\end{table}

The experiments were run on a 2.9 GHz Intel Core i5 16 GB RAM computer. We configured the cloud node with QoS requirements as defined in Table \ref{tbl_QoSAttributes}, tactics as defined in Table \ref{tbl_QualityTactics}, and adaptation rules as defined in Table \ref{tbl_AdaptationRules}. To examine the accuracy of simulation results, we examined quality attributes at each time interval of 864 seconds in the cases of self-adaptive, stimulus-aware, goal-aware and non-adaptive architectures, i.e. we run the entire workload for each service type and measured the quality attributes. 

\subsection{Validation Results}
\begin{figure*}[!h]
\centering
\includegraphics[width=\textwidth]{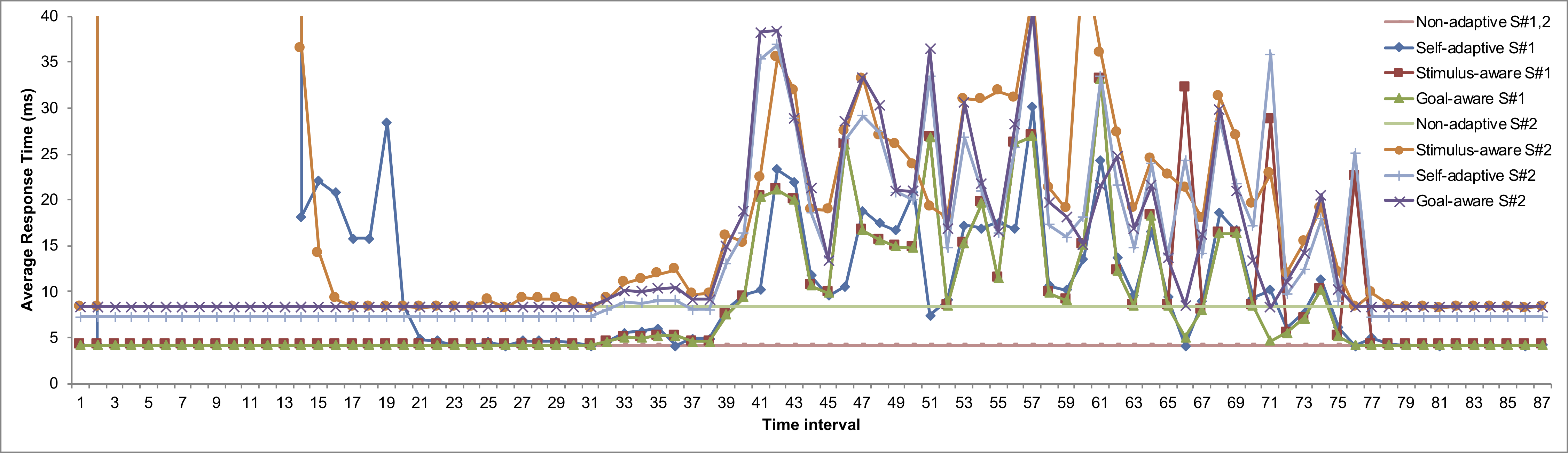}
\caption{Average response time of service types 1 and 2 in time intervals}
\label{graph_resultsIntervalsResponseTime}
\end{figure*}

\begin{figure*}[!h]
\centering
\includegraphics[width=\textwidth]{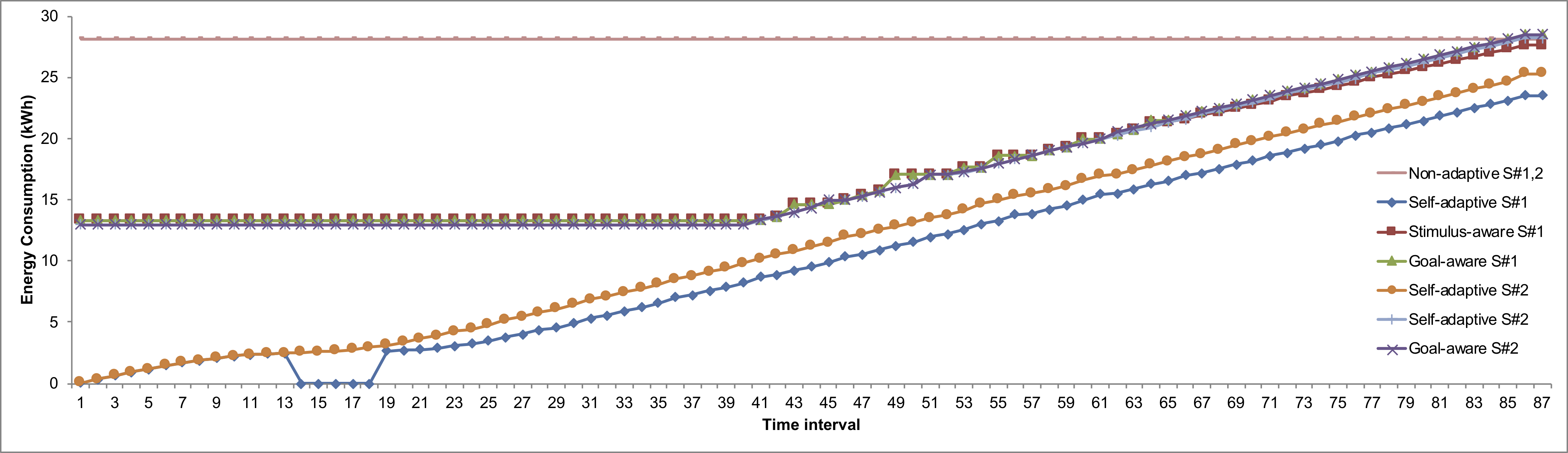}
\caption{Average energy consumption of service types 1 and 2 in time intervals}
\label{graph_resultsIntervalsEnergy}
\end{figure*}

\begin{figure*}[!h]
\centering
\includegraphics[width=\textwidth]{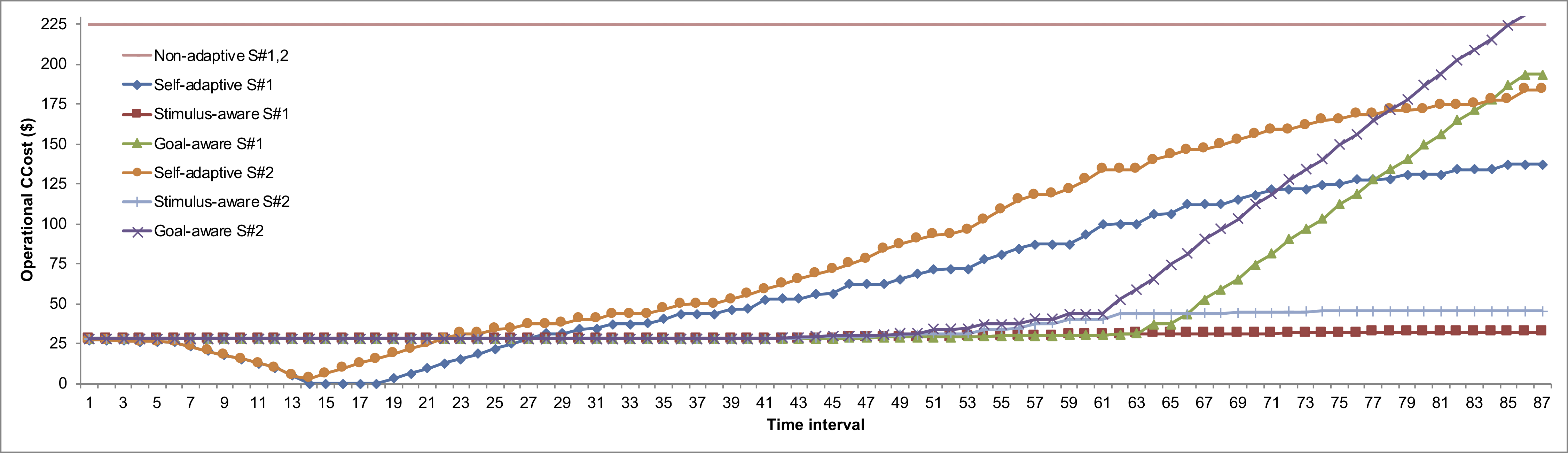}
\caption{Average operational cost of Service Types 1 and 2 in time intervals}
\label{graph_resultsIntervalsCost}
\end{figure*}

To validate the simulation environment, we compare the average response time, operational cost and energy consumption of all architectures during the experiment time intervals, as shown in Figures \ref{graph_resultsIntervalsResponseTime}, \ref{graph_resultsIntervalsEnergy} and \ref{graph_resultsIntervalsCost} respectively for service types 1 and 2 (with the least and most processing requirements). As the non-adaptive architecture was running on a static configuration (same number of hosts and VMs required to handle the highest load), the results of response time are the same for time intervals. The adaptive and aware architectures have similar values like the non-adaptive architecture during off-peak intervals, where they were able to handle the workload without adaptations. During peak intervals, response time started to fluctuate, where adaptations took place to meet the goal. As expected, the operational cost and energy consumption of the latter architectures are lower than the non-adaptive architecture, with a maximum equal to the values of the non-adaptive architecture. These are the expected behaviours for all architectures considering the testbed configurations. Hence, the results reflected that architectures components are correctly implemented. Obviously, the results showed the benefits of adaptivity and awareness with respect to achieve required performance, while saving operational cost and energy consumption. 

\subsection{Performance Evaluation}
In order to evaluate the performance of self-adaptive and self-aware architectures, we observe a closer look at the processing of all service requests and compared the percentage of response time violations for different service types. As expected, the goal-aware architectures had the less violation percentage (e.g. 24.40\% in the case of service type 2). This is due to the proactive adaptation taken prior to violations. While the self-adaptive had better performance than stimulus-aware (e.g. 26.44\% versus 28.86\% in the case of service 2), operational cost was remarkably higher in the former case starting from the peak time (as shown in Figure \ref{graph_resultsIntervalsCost}).

\begin{figure}[!h]
\centering
\includegraphics[width=0.75\textwidth]{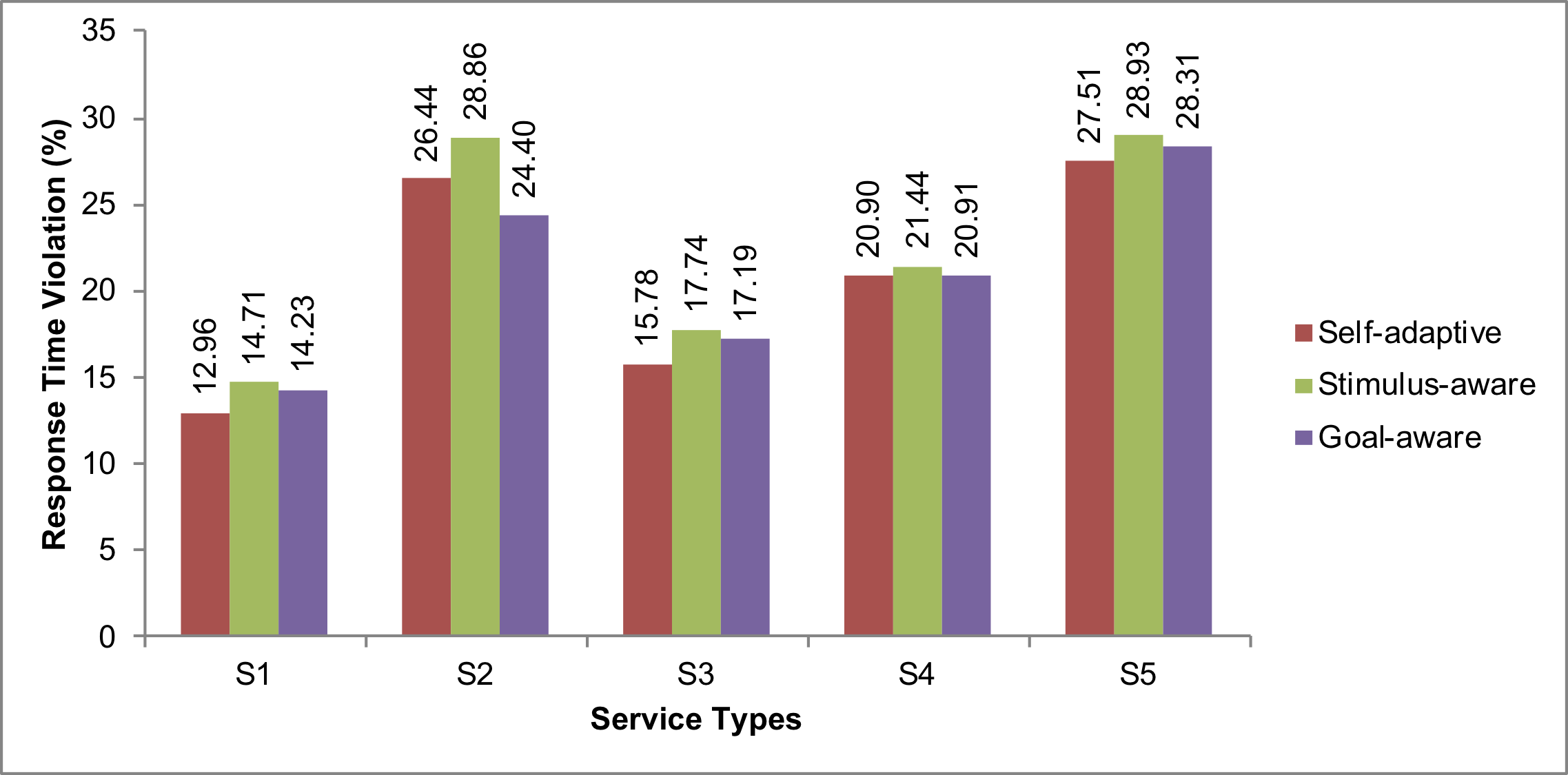}
\caption{Response time violations (\%)}
\label{grapgh_resultsResponseTimeViolations}
\end{figure}

Considering the experiments total results, we report the average results of the whole experiment for each service type and the average of each architecture in Table \ref{tbl_resultsQualityAttributes}. The non-adaptive architecture had a fixed value for all attributes, due to the static configuration. The average response time of all requests for each service type is much better achieved by the goal-aware architecture due to proactive adaptations, followed by stimulus-aware and self-adaptive architecture (average 20.02, 20.53, 62.85 ms respectively). While achieving better performance, energy consumption (calculated based on the number of running hosts) and operational cost (calculated based on the number of running VMs) were found less on average than non-adaptive. For instance, average energy consumption is 17.42, 17.32, 11.37 kWh versus 28.14 kWh for the non-adaptive architecture, due to consolidation performed during off-peak periods and scaling during peak load only. Operational cost is found less in the case of stimulus-aware architecture (31.88 \$), followed by the goal-aware (56.26 \$) and self-adaptive (79.57 \$) compared to non-adaptive (224.34 \$). As the stimulus- and goal-aware architectures were running nearly the same number of hosts, their energy consumption were close. But, each was running different number of VMs, which caused the difference in operational cost. The goal-aware architecture used a higher number of VMs in pro-active adaptations.

\begin{table*}[!h]
\caption{Experiments average results}
\label{tbl_resultsQualityAttributes}
\center
\footnotesize
\begin{tabular}
		{	
		>{\raggedright}p{0.15\textwidth} 
		l
		l l l l
		}
\toprule								
\textbf{Experiments results	}		&
{\textbf{S\#}}									&
\multicolumn{4}{c}{\textbf{Architecture}} 
\\
										&
										&
Non-adaptive				&								
Self-adaptive				&	
Stimulus-aware			&
Goal-aware
\\
\midrule
\multirow{6}{*}{\parbox{2cm}{Response Time (ms)}}		 		&	
1								&	
4.17							&
73.73						&	
16.54						&
16.00
\\
								&	
2								&
8.33						&
63.49						&	
23.01						&
22.92
\\
								&	
3								&	
5.00						&
58.41						&
19.18						&	
18.56
\\
								&	
4								&	
6.25						&
58.90						&
21.69						&	
21.10
\\
			 					&	
5								&	
7.08						&
59.74						&	
22.24						&
21.54
\\
										&
avg.								&
6.17									&
62.85								&
20.53								&
20.02
\\
\hline
\multirow{6}{*}{\parbox{2cm}{Average energy consumption (kWh)}}				&	
1										&	
28.14								&	
10.42								&
17.42								&
17.56
\\
			 							&	
2										&	
28.14								&
11.61								&
17.30								&	
17.37
\\
									&	
3									&
28.14							&
11.61							&
17.35							&	
17.41
\\
									&	
4									&	
28.14							&
11.61							&
17.29							&	
17.39
\\
									&	
5									&
28.14							&	
11.61							&
17.27							&	
17.36
\\
										&
avg.								&
28.14								&
11.37								&
17.32								&
17.42
\\
\hline
\multirow{6}{*}{\parbox{1.8cm}{Total operational cost (\$)}}			 							&	
1									&	
224.34						&
64.54							&
29.36							&	
52.40
\\
								&	
2								&
224.34					&
84.41						&
34.08						&	
64.47
\\
								&	
3								&
224.34					&	
82.61						&
30.43						&	
53.68
\\
								&	
4								&	
224.34					&
82.61						&
31.65						&	
56.29
\\
										&
avg.								&
224.34							&
79.57								&
31.88								&
56.26
\\
\bottomrule
\end{tabular}
\end{table*}

\subsection{Evaluation of Adaptation Overhead}
\begin{figure}[!h]
\centering
\includegraphics[width=0.75\textwidth]{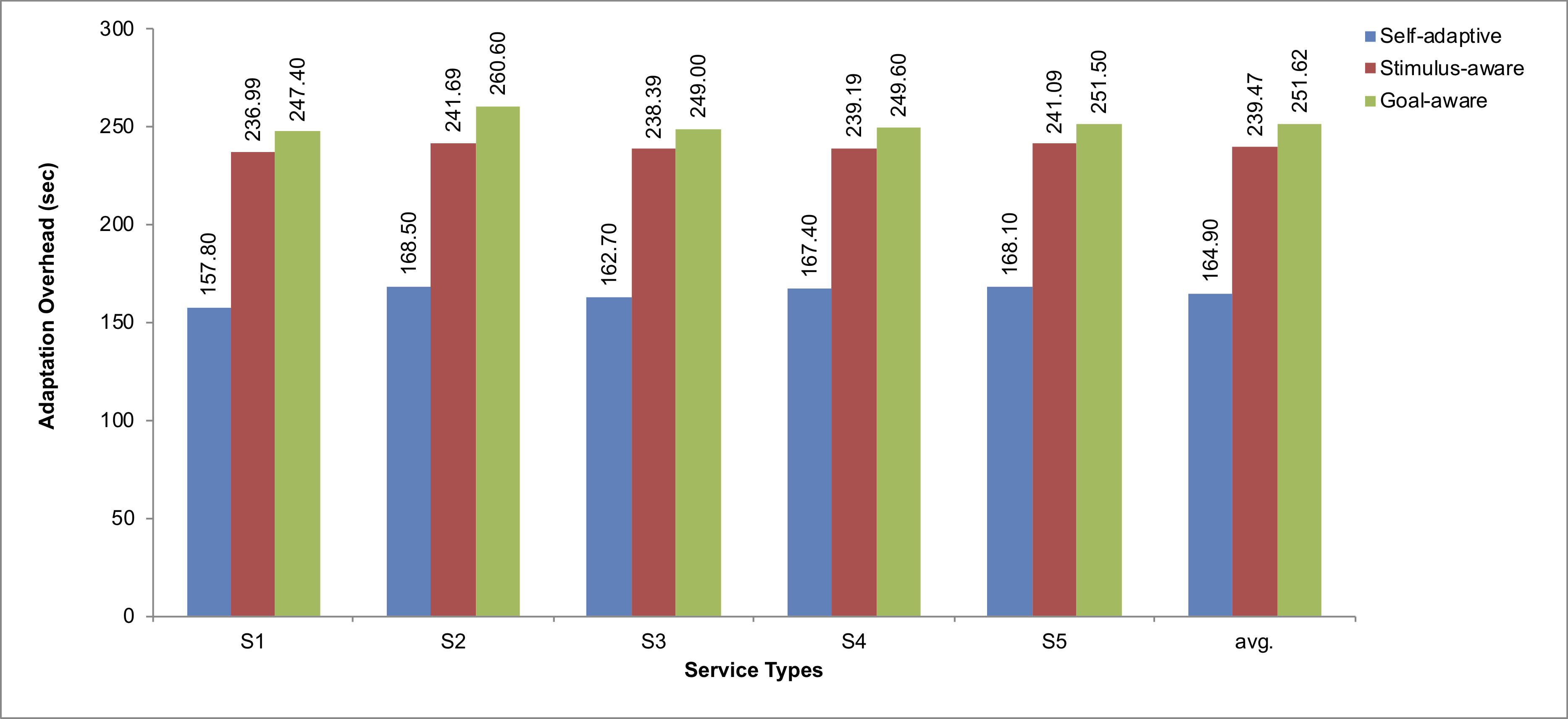}
\caption{Adaptation Overhead (sec)}
\label{graph_resultsAdaptationOverhead}
\end{figure}

We evaluate the adaptation overhead by calculating the total time spent by the architecture in monitoring quality attributes, detecting violations, making and executing adaptation decisions. Figure \ref{graph_resultsAdaptationOverhead} shows the overhead of each service type and their average. As goal-aware architecture is performing pro-active adaptations, its overhead is the highest (251.62 sec on average). Stimulus-aware is close to goal-aware due to the intelligent reactions (239.47 sec). The overhead of self-adaptive is lower (164.90 sec) due to reactive adaptations, which obviously resulted in lower performance.

\subsection{Additional Use Cases for Evaluation}
The framework was used in our earlier works: (i) to evaluate runtime workload requests assignments, dynamic scheduling policies and queuing models \cite{Salama2016} \cite{Salama2018b}, (ii) in modelling and testing tactics impact on the stability of self-adaptive and self-aware architectures \cite{Salama2016}, (iii) modelling different self-aware and self-adaptive architecture patterns \cite{Salama2018b}. We are currently employing the framework as a symbiotic simulator for behavioural stability in running what-if-scenarios and compare the quality of adaptation.

\section{Threats to Validity}
\label{sec_threats}
There are some potential threats to validity of the proposed work:

\begin{itemize}
\item The fact that the proposed work is evaluated by its authors presents a threat to objectivity. To mitigate this risk, we sought to conduct other sets of experiments with different testbeds, in order to ensure the feasibility of the toolkit. We, also, plan to conduct other evaluation case studied in other research contexts.

\item Subjectivity might be considered a threat to validity in setting the QoS attributes, as it was conducted based on the authors' background and knowledge. Our mitigation strategy for this issue has been to base the case study on previous work of \cite{Chen2014a} \cite{Salama2015} \cite{Salama2016} \cite{Salama2017}, this makes us believe that the evaluation setup is practical and challenging.

\item Experiments were conducted in a controlled environment and have not considered the real-life scenario of switching between different service patterns and changing user requirements during runtime for different end-users. Given the use of a real-world workload trend and the \textit{RUBiS} benchmark, we consider that our experiments have given good enough indication and approximation of likely scenarios in a practical setting. Also, we have chosen the QoS goals thresholds purely based on our observations, e.g. response time not exceeding 25 ms. Yet, these goals have proved to be challenging when running the experiments. 

\end{itemize}

\section{Conclusion and Future Work}
\label{sec_conclusion}
In this paper, we proposed \textit{SAd/SAw-CloudSim}, a modelling and simulation environment for self-adaptive and self-aware cloud architectures, extending \textit{CloudSim} with novel extensions useful for modelling and testing self-adaptivity and self-awareness. The toolkit allows running dynamic runtime workload, and can be used as a symbiotic simulator during runtime. Our future work focuses on testing the simulation environment in new research contexts, other benchmarks and quality requirements. We also aim to extend it for managing scientific workflows. The next development will include the implementation of interaction-awareness in the context of cloud federations and geo-distributed cloud datacenters.

\section*{Acknowledgments}
Authors would like to specially thank Maria A. Rodriguez for the support in developing and testing the toolkits. This research was supported in part by the Universitas 21.

\section*{References}
\bibliography{SAwCloudSim-bib}

\end{document}